\documentclass[prl,aps,twocolumn, floatfix]{revtex4}
\usepackage{amssymb}
\usepackage{graphicx}
\usepackage{dcolumn}
\usepackage{bm,amsmath,verbatim}

\def\be{\begin{equation}}
\def\ee{\end{equation}}
\def\bea{\begin{eqnarray}}
\def\eea{\end{eqnarray}}
\def\bse{\begin{subequations}}
\def\ese{\end{subequations}}

\def\be{\begin{eqnarray}}
\def\ee{\end{eqnarray}}

\newcommand{\ua}{\uparrow}
\newcommand{\da}{\downarrow}

\newcommand{\expect}[1]{\langle {#1} \rangle}

\begin{document}
\title{On the possibility of the fractional ac Josephson effect in non-topological conventional superconductor-normal-superconductor
 junctions}
\author{Jay D. Sau}
\author{Erez Berg}
\author{Bertrand I. Halperin}
\affiliation{ Department of Physics, Harvard University, Cambridge, MA, 02138, USA}
\date{\today}
\begin{abstract}
Topological superconductors supporting Majorana Fermions with
non-abelian statistics are presently a subject of intense
theoretical and experimental effort. It has been proposed that the
observation of a half-frequency or a fractional Josephson effect
is a more reliable test for topological superconductivity than the
search for end zero modes. 
Low-energy end modes can occur accidentally due to impurities.
In fact, the fractional Josephson effect has been observed for the semiconductor nanowire system.
 Here we consider the ac Josephson
effect in a conventional $s$-wave superconductor-normal
metal-superconductor junction at a finite voltage. Using a
Floquet-Keldysh treatment of the finite voltage junction, we show
that the power dissipated from the junction, which measures the ac
Josephson effect, can show a peak at half (or even incommensurate fractions) of the Josephson frequency.
 The ac fractional Josephson peak can also be understood
simply in terms of Landau-Zener processes associated with the Andreev bound state spectrum of the junction.
\end{abstract}

\maketitle

\paragraph{Introduction}
Topological superconductors \cite{schnyder} are promising candidates for
the practical solid state realization of Majorana Fermions (MF) \cite{p_wave_sf,yakovenko1,kitaev,
chuanwei,fu_prl'08,sau,long-PRB,alicea,roman,oreg}.
The MFs are predicted to occur as zero-energy bound states attached to defects and have received much attention recently
~\cite{Marchmeeting,Wilczek-3}, both due to their predicted non-Abelian statistics \cite{alicea1,david,bert,beenakker1}
and their potential application in topological quantum computation
 (TQC)\cite{nayak_RevModPhys'08,Wilczek-3,Moore,Kitaev,Wilczek2}.
 A simple topological superconducting (TS)
system supporting MFs, which has attracted considerable experimental attention \cite{Wilczek-3}, consists of a
 spin-orbit coupled semiconducting system in a magnetic field placed in contact with an ordinary superconductor
 \cite{sau,long-PRB,alicea,roman,oreg}.
MFs at the ends of such a wire have been predicted to produce a zero-bias conductance peak~\cite{long-PRB,yakovenko1,ZBCP}.
 In fact, recent experiments \cite{kouwenhoven_science,heiblum,xu_h_q} measuring the tunneling conductance
in the semiconducting wire system suggest the existence of MFs.
\begin{figure}[!h]
\centering
\includegraphics[scale=0.4,angle=270]{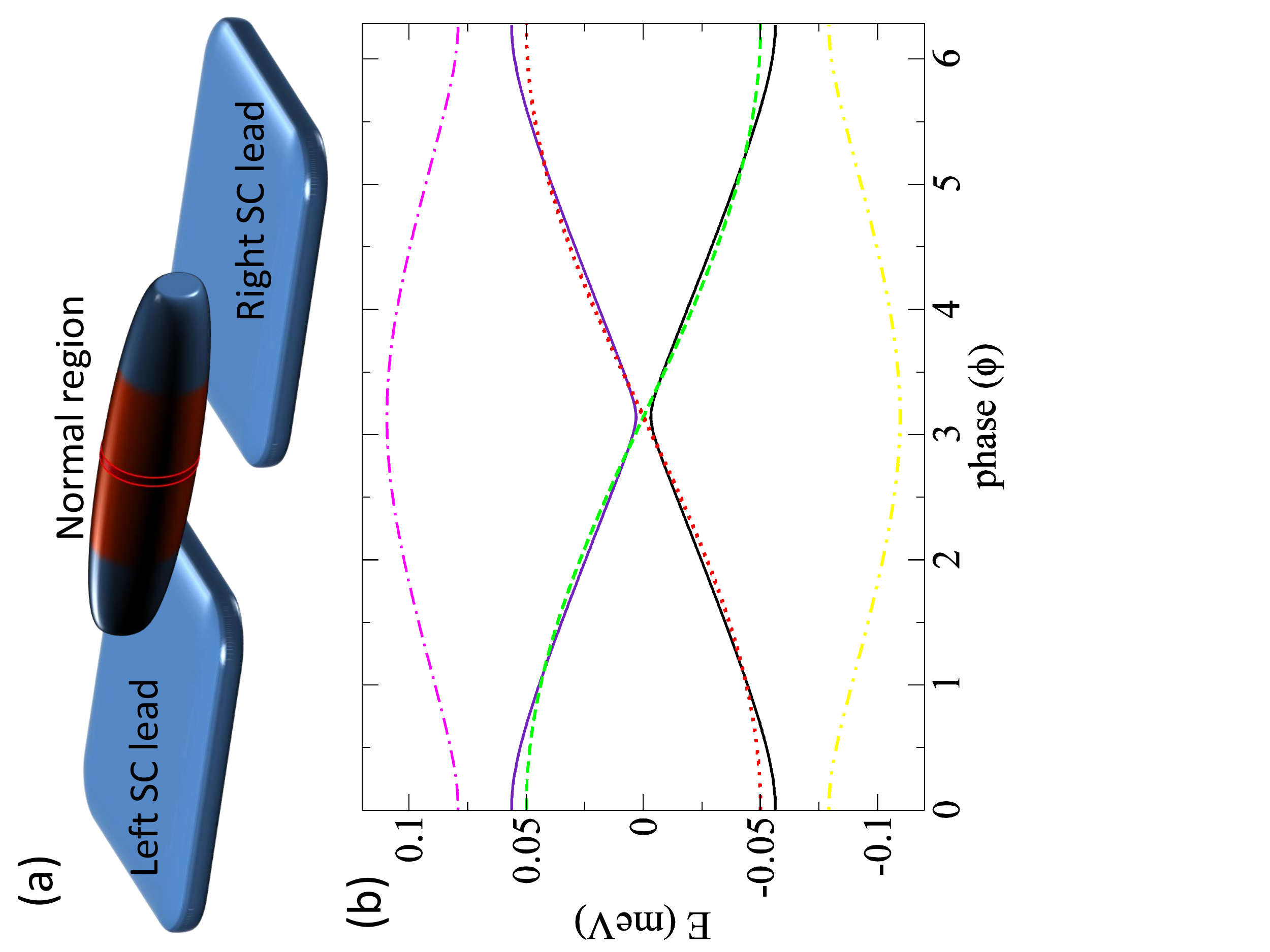}
\caption{
(Color online)(a)Low-density normal region with chemical potential $\mu$ in between
 two $s$-wave superconducting (SC) leads with phase difference $\phi$.
The normal region is long enough to support confined ABSs with energy less than the SC gap $\Delta_0$ in the
SC leads. (b)ABS energy ($E$) spectrum in the normal region as a function of $\phi$ for $\Delta = 0.4$ meV and $\mu = 0.13$ meV. The ABSs in the
conventional superconducting junction (solid and dash-dotted lines) show a weak avoided crossing
 at $\phi=\pi$\cite{beenakker,averin}, while the ABSs in
 the topological superconducting states (dashed and dotted lines) show a protected crossing \cite{kitaev,yakovenko,roman,oreg}.
}\label{Fig1}
\end{figure}

While the zero-bias conductance peak  observations \cite{kouwenhoven_science,heiblum,xu_h_q} are encouraging,
peaks in the tunneling spectrum arising from impurities are difficult to completely rule out.
The end MFs appear in the TS system as a result of a Pfaffian topological invariant associated with the bulk ~\cite{kitaev,parag,schnyder}
of the wire. In a TS system in the ring geometry, with a weak-link (shown in Fig.~\ref{Fig1}(a)) connecting the ends,
 the  bulk topological invariant leads to
topologically protected crossings of localized Andreev bound
states (ABS) as a function of the flux penetrating the ring 
\cite{kitaev,roman}.
The crossings of the ABS of an SNS junction can, in principle, be inferred  using the ac Josephson
effect with a finite DC voltage $V$ across the junction so that the phase varies in time as $\phi(t)=2\pi\frac{\Phi}{\Phi_0}=\Omega_J t$.
Here $\Omega_J=2 e V$ is the Josephson frequency, where we have set $\hbar=1$.
If the applied voltage $V$ is small enough, the time-evolution of the junction can
be considered to be adiabatically following the ground state of the SNS junction with a phase difference $\phi$.
The current in the junction $I(t)=V^{-1}\frac{d E}{d t}=\frac{d E}{d \phi}$ (where $E$ is the total energy)
 then varies in time corresponding to the ABS energies.
If a pair of ABSs (shown as dashed lines in Fig.~\ref{Fig1}(b))
 cross zero energy, evolution of the state following the ABSs as $\phi$ varies from $0$ to $2\pi$  leads to a state with a different energy 
at the end of each period $t=\tau=2\pi/\Omega_J$.
This  results in a
current-phase relation with a component $I(\phi)\sim \sin{\frac{\phi}{2}}$,
 which is at double the period of the conventional Josephson effect \cite{kitaev,yakovenko,fu_kane,roman,oreg,rosenow},
 and is referred to as the fractional
Josephson effect.
  Here $\phi$ is the gauge invariant phase across the superconducting tunnel junction and is given by $\phi=2\pi\Phi/\Phi_0$,
where $\Phi$ is the magnetic flux through the loop and $\Phi_0=hc/2e$. However, at finite temperature or in the presence of a 
finite density of fermionic states, the excited ABS state for $\phi>\pi$ can relax to the lower energy state so that fractional Josephson 
effect in TS systems can be observed only above a finite frequency \cite{fu_kane}.
For conventional superconductors without such zero-energy crossings of the ABSs (solid lines in Fig.~\ref{Fig1} (b)),
the current in the adiabatic and low-temperature limit is $2\pi$-periodic.

The periodically varying current $I(t)$, for a
junction at constant voltage $V$, will lead to radiation with a power
spectrum $P(\omega) \propto <|I(\omega)|^2> $
 with peaks corresponding to the Fourier components
of the current--phase relation $I(\phi)$. TS systems in such
configurations have been predicted  to show peaks at half the
Josephson frequency i.e. at $\omega=\Omega_J/2$. The
power-spectrum radiated from junctions is somewhat difficult to
measure and it is more convenient to look for absorption of
microwaves by an SNS junction leading voltage steps called Shapiro
steps. The Shapiro steps occur at voltages where the Josephson
frequencies of the corresponding voltage matches the applied
frequency~\cite{tinkham}.
 Indeed, the fractional current-phase relationship has been observed in the semiconductor nanowire system in the form of   double
 voltage Shapiro steps, strengthening the
evidence for the topological superconductivity in these
systems~\cite{rokhinson}.

 In this paper, we show that, even non-topologial superconducting systems, such as a conventional SNS junction constructed
out of $s$-wave superconductors can produce an ac fractional
Josephson (or Shapiro) effect in $P(\omega)$, if the applied
voltage $V$ happens to be larger than an avoided crossing gap
(shown in Fig.~\ref{Fig1}) that can accidentally become small,
 even though the current-phase relation of such a system would be $2\pi$ periodic.
Thus, similar to zero-bias tunneling signature for MFs, the fractional Josephson effect can also occur accidentally in
conventional systems.
For our study, we will ignore the effects of Coulomb interactions. In the high transparency parameter regime that 
we study one can expect these effects to be small and only effectively renormalize the parameters such as position of the 
crossing of ABSs as a function of chemical potential. The conventional SNS junction only serves as a model to study the fractional Josephson effect.
Similar physics is expected to hold in other systems where pairs of ABSs can cross zero-energy as a result of tuning of parameters. 

 To see how an SNS junction can give rise to a fractional ac Josephson effect, we consider the Hamiltonian
 of an $s$-wave SNS junction shown in Fig.~\ref{Fig1}(a)
under a finite voltage  $V$, which is written as $\hat{h}(\phi(t))=\int
dx \Psi^\dagger(x) h[\phi(t)]\Psi(x)$ where
$\Psi(x)=(\psi_{\ua}(x),\psi_{\da}^\dagger(x))$ is the fermion
operator in Nambu spinor notation and $\phi(t)=\Omega_J t$
 is the time-dependent phase difference between the left and the
right superconducting contacts. Here $h$ is the Bogoliubov-de
Gennes Hamiltonian, given by
\begin{equation}
h[\phi]=[(-\nabla^2-\mu)\tau_z+\Delta_0\{(\Theta(x+W)+\Theta(W-x)e^{i \phi})\tau_++h.c\}],\label{eq:H}
\end{equation}
where $\tau_{z,+,-}$
are the corresponding particle-hole Pauli matrices. In Eq.~\ref{eq:H},  $\Delta_0$ is the superconducting gap in the S region, $2 W$
is the width of the junction, $\mu$ is the chemical potential, which controls the density of electrons in the normal region and
 $\Theta(x)=1$ for $x>0$ and $\Theta(x)=0$ for $x<0$.
Such SNS junctions have been shown to have ABS spectra with avoided crossings controlled by the transparency of the
barrier \cite{beenakker}. It has also been shown that Landau-Zener (LZ) crossing processes can play an important role
in determining the dc current response ~\cite{averin} in conventional SNS and the finite-frequency response \cite{fu_kane}
in TS junctions.
 In the case of a high transparency interface with
 reduced Fermi-energy in the N region leads to an ABS spectrum shown in Fig.~\ref{Fig1}
 (solid lines) with a gap near $\phi=\pi$, which is controlled by  the chemical potential $\mu$.
The phase-dependent
super-current operator $\hat{I}(\phi)$ is obtained as a derivative
\begin{equation}
\hat{I}(\phi)=\partial_\phi \hat{h}[\phi].\label{eq:I}
\end{equation}

 For conventional SNS junctions, one expects the ABS occupation to stay in the ground state
 for voltages $2 e V\ll E_g$, where $E_g$ is the minimal gap in the junction. However, typical high-transparency and near resonant
SNS junctions show only weakly avoided crossings (as seen in Fig.~\ref{Fig1}), so that
so that 2 eV can be equal or larger  than $E_g$ leading to a
violation of the adiabaticity requirement. 
In this case, the state of the system can crossover from one state to the other by LZ tunneling leading to a situation analogous
to the ABS in the TS case (shown by dotted lines in Fig.~\ref{Fig1}), resulting in a  fractional ac Josephson effect even
in the conventional SNS junction. For simplicity, we consider values of $\mu$ and $W$, where the gap in the spectrum (between the lowest positive 
energy excitations) near
 $\phi\sim 0$ is much larger than the energy of the lowest energy excitation near $\phi=\pi$
 (as in solid lines in Fig.~\ref{Fig1}(b)). In this parameter regime, a range of voltages $V$ allow the dynamics of the lowest 
energy states to remain confined to the lowest energy pair of states.

The power-spectrum $P(\omega)$ dissipated by the fluctuating current $\hat{I}(t)$  in the
Josephson junctions with a weak applied DC voltage $V$, can be
calculated \cite{mahan}  using the expression 
\begin{align}
&P(\omega)= P_c(\omega)+\sum_n |\tilde{I}_n|^2\delta(\omega \tau-2\pi n),\label{eq:P}
\end{align}
where $P_c(\omega)$ is the connected part of the power spectrum  defined as 
\begin{align}
&P_c(\omega)\sim \int_0^\tau \int_{-\infty}^{\infty} \frac{dt_1 dt_2}{\tau}e^{-i\omega(t_1-t_2)}\expect{\hat{I}(t_1)\hat{I}(t_2)}_c,\label{eq:Pc0}
\end{align}
with $\expect{\hat{I}(t_1)\hat{I}(t_2)}_c=\expect{\hat{I}(t_1)\hat{I}(t_2)}_c-\expect{\hat{I}(t_1)}\expect{\hat{I}(t_2)}$
and 
the fourier transform $\tilde{I}_n$ is defined as 
\begin{equation}
\tilde{I}_n=\int_0^\tau dt e^{i n\Omega_J t}\expect{\hat{I}(t)}\label{eq:In}.
\end{equation}
Here we have generalized the expression appropritately to include discrete time-translation invariance
of the Hamiltonian (i.e. $t\rightarrow t+\tau$). The  time-periodic 
expectation value of the current in Eq.~\ref{eq:In} gives rise to singularities in the power-spectrum in Eq.~\ref{eq:P}.

For simplicity, let us first consider the case where the current
operator $\hat{I}(t)$ is strongly coupled to ohmic dissipation
from a resistive shunt so that quantum fluctuations of the
current $\hat{I}(t)$ are suppressed. The current operator 
$\hat{I}(t)$ can then be replaced by a classical random variable $\hat{I}(t)\rightarrow \bar{I}(t)$.
In this classical limit, the current $\bar{I}(t)$ depends on the
quasiparticle occupation of the SNS junction according to
\begin{equation}
\bar{I}(t)=s(t)i_0(t),\label{eq:markov}
\end{equation}
 where $i_0(t)$ is the current of
the ABS in the lower energy state
and $s(t)$ is a random variable such that 
$s(t)=-1$ in the higher energy state in Fig.~\ref{Fig1}, while
$s(t)=1$ in the lower-energy state. For now, we have assumed that
the fermion parity in the junction is fixed, so that the junction
has only two states. The state variable $s(t)$ is assumed to be
constant, except near the avoided crossings where $\phi(t)=2 e V t$
crosses $(2 n+1)\pi$, where an LZ crossing can transfer $s(t)$
between the values $\pm 1$ with some probability. The evolution of
$s(t)$  can then be described by a classical Markov process with a
transition probability matrix $P\left[s(t=n\tau)\rightarrow
s(t=(n+1)\tau)\right]$. Using the transition probabilities, the connected part of 
the power spectrum for the current fluctuations according to Eq.~\ref{eq:Pc0} is found to be
\begin{align}
&P_c(\omega)=\frac{4 a b (2-a-b)}{(a+b)}\frac{1}{(a+b-1+\cos{\omega\tau})^2+\sin^2{\omega\tau}}\nonumber\\
&(\sum_m  \tilde{i}_m \frac{\sin{\omega\tau/2}}{\omega-m e V})^2\label{eq:mc},
\end{align}
where $i_0(t)=\sum_m \tilde{i}_m e^{i 2 m e V t}$ and $a,b$ are the conditional transition probabilities at each
LZ crossing at time $t=n\tau$  so that $a=P\left[s(t=(n+1)\tau)=-1|s(t=n\tau)=+1\right]$ and $b=P\left[s(t=(n+1)\tau)=+1|s(t=n\tau)=-1\right]$.
 At high voltages $V$, which are much higher than the avoided crossing, where the LZ crossing probabilities $a\sim b\sim 1$, it
 follows from Eq.~\ref{eq:mc} that the radiated power $P(\omega)$ has a peak(from the vanishing denominator) at $\omega\sim\pi/\tau=\Omega_J/2$
that is characteristic of the fractional ac Josephson effect.
In addition, for asymmetric transition probabilities $a\neq b$ the power-spectrum in Eq.~\ref{eq:P} contains 
a train of singularities at $\omega=2\pi/n$ with strength
\begin{equation}
|I_n|^2=\frac{(b-a)^2}{(a+b)^2}|\tilde{i}_n|^2.
\end{equation}

For large voltages $V$, one expects the state of the junction to become highly excited and possibly change fermion number while remaining
 quantum coherent over a few periods $\tau$.
 Therefore, we consider an SNS junction described
by the time-dependent  BCS Hamiltonian $\hat{h}[\phi(t)]$ (Eq.~\ref{eq:H}), while being weakly tunnel-coupled
 to a fermionic bath.
The Keldysh time-contour Green-function \cite{mahan, baym,kamenev} for the system including the time-dependent Hamiltonian $h[\phi(t)]$
is written in the form of a Dyson equation
\begin{equation}
G=g+g h[\phi(t)]G,\label{dyson}
\end{equation}
where $g$ is the Keldysh time-contour Green-function for the system excluding the time-dependent part of the Hamiltonian(i.e. $h[\phi(t)]$).
The Green function $g$ is thus an equilibrium Green function is written as
\begin{align}
&g_{R,A}(\omega)=\frac{1}{\omega\pm i\Gamma},\,\quad g^{<,>}(\omega)=\frac{\pm 2 i\Gamma}{\omega^2+\Gamma^2}n_F(\pm\omega),\label{eq:g}
\end{align}
where $\Gamma$ is the imaginary part of the self-energy (assumed to be independent of $\omega$) arising from coupling to the fermionic bath
and $n_F(\omega)=\frac{1}{1+e^{(\omega/T)}}$ is the fermion distribution function~\cite{mahan,baym}.
Using the Green function in Eq.~\ref{eq:g} in the Dyson equation Eq.~\ref{dyson} one finds the retarded Green function to be
\begin{equation}
G_{R,A}(t,t')=\mp i\vartheta(\pm(t-t'))U(t,t'),\label{eq:GR1}
\end{equation}
 where   $\vartheta(t)=e^{-\Gamma t}\Theta(t)$, and $U(t,t')=\mathcal{T} e^{-i\int_{t'}^t dt_1 h(t_1)}$.
Expanding out the Dyson equation Eq.~\ref{dyson}~\cite{mahan}, and
using the relations $U(t_1,t_2)h(t_2)=-i\partial_{t_2}U(t_1,t_2)$ together with  $h(t_1)U(t_1,t_2)=i\partial_{t_1}U(t_1,t_2)$,
the distribution functions are found to be
\begin{equation}
G^{<,>}=\mp 2 i G_R \tilde{n}_F(\pm(t-t')) G_A\label{GK},
\end{equation}
where $\tilde{n}_F(t)$ is the Fourier transform of $n_F(\omega)$
In the $T\gg V$ limit, we can approximate $\tilde{n}_F\sim\delta(t-t')$ and write
\begin{equation}
G^{<,>}(t,t')=\pm 2 i e^{-\Gamma|t-t'|}U(t,t').\label{Gl}
\end{equation}
Using the Floquet theorem~\cite{floquet_GF}, the unitary time-evolution operator
 $U(t_1,t_2)$ corresponding to the time-periodic Hamiltonian $h[\phi(t)]$
can be constructed in terms of Floquet states $\varphi_{\lambda}(t)$.
The Floquet states $\varphi_{\lambda}(t)$ are defined at $t=0$ to be eigenstates of $U(\tau,0)$ 
using the relation  $U(\tau,0)\varphi_{\lambda}(0)=e^{i\epsilon_\lambda\tau}\varphi_\lambda(0)$. Here $\epsilon_{\lambda}$ 
are the Floquet energies corresponding to $\varphi_{\lambda}(t)$. The definition of $\varphi_{\lambda}(t)$ is then extended to all time using the 
relation $\varphi_\lambda(t+\tau)=\varphi_\lambda(t)=e^{-i\epsilon_\lambda t}U(t,0)\varphi_\lambda(0)$. 
The unitary operator $U(t,0)$, which is required for calculating the Floquet states, 
 can be computed numerically by solving the time-dependent Schrodinger equation corresponding to
Eq.~\ref{eq:H}.
The unitary time-evolution operator can then be conveniently written in the basis of the Floquet states as 
\begin{align}
&U(t_1,t_2)=\sum_\lambda e^{-i\epsilon_\lambda (t_1-t_2)}\varphi_\lambda(t_1)\varphi_\lambda^\dagger(t_2).\label{eq:Uf}
\end{align}

Restricting to the two ABS state case, the Floquet energies, $\epsilon_\lambda$, are determined from the eigenvalues $e^{i\epsilon_\lambda\tau}$
of the unitary matrix $U(\tau,0)$.
Since the underlying BCS Hamiltonian in Eq.~\ref{eq:H} is particle-hole symmetric, $U(\tau,0)$
is restricted to the form
\begin{equation}
U(\tau,0)=\left(\begin{array}{cc}\cos{\theta}e^{i\phi_0}&-\sin{\theta}e^{-i\phi_1}\\\sin{\theta}e^{i\phi_1}&\cos{\theta}e^{-i\phi_0}\end{array}\right)
\end{equation}
so that
\begin{equation}
\epsilon_{\lambda}=\pm\frac{2 e V}{2\pi}\textrm{sgn}(\cos{\phi_0}) \cos^{-1}{\{\cos{\theta}\cos{\phi_0}\}}.\label{eq:epslambda}
\end{equation}

\begin{figure}
\centering
\includegraphics[scale=0.35,angle=0]{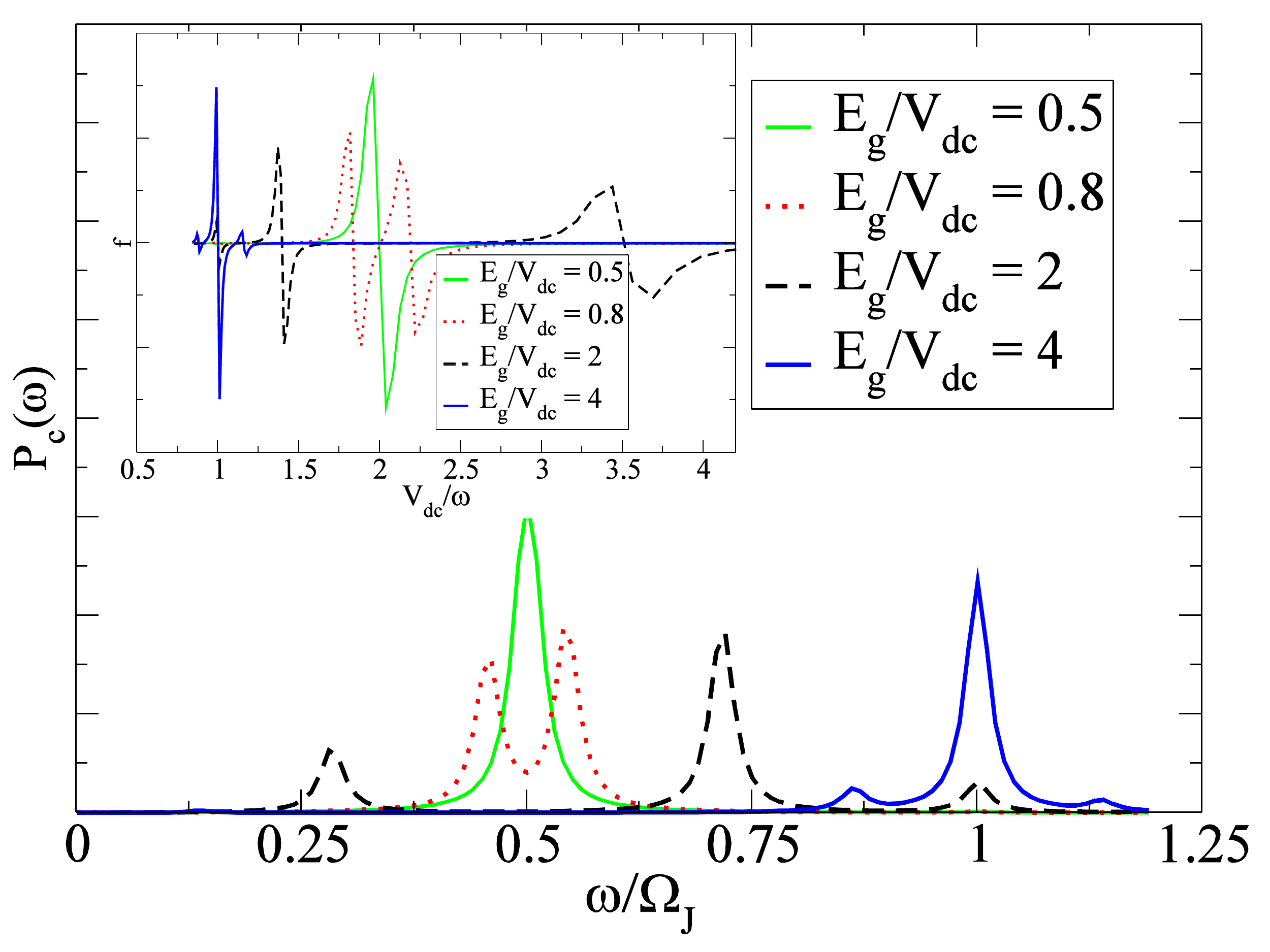}
\caption{
Power dissipated $P_c(\omega)$ as a function of frequency $\omega/\Omega_J$ for
 different values  of the ratio of the Andreev gap $E_g$ (i.e. half the avoided crossing in the
 ABS near $\phi=\pi$) to the applied voltage $V_{dc}$.
The ratio $E_g/V_{dc}$ is varied by changing $\mu$.  The power-spectrum $P_c$, for small $E_g/V_{dc}$, shows
 a peak at $\omega=\Omega_J/2$. In contrst, $P_c(\omega)$ shows only the conventional Josephson peak at $\omega= \Omega_J$ for 
larger values of $E_g/V_{dc}$.
Here $P_c(\omega)$ is only the connected part of the radiated power spectrum (see Eq.~\ref{eq:P}). The total power $P(\omega)$ may contain
additional contributions at integer frequencies $\omega\sim n\Omega_J$. The inset shows the Shapiro kink 
response $f(V_{dc})=(I_{tot}R_1-V_{dc})/R_1$ of the DC voltage  $V_{dc}$
as the bias current $I_{tot}$ changes.
The function $f$ shows peaks at both the conventional voltage $V_{dc}=\hbar\omega/2 e$ and the doubled Shapiro
 voltage $V_{dc}=2\hbar\omega/2 e$ as a function of the ratio $V_{dc}/\omega$.
 Here $V_{dc}=1\,\mu$eV and $\mu$ ranges from $0.12$ meV to $0.15$ meV.
}\label{Fig2}
\end{figure}

The connected correlator $\expect{\hat{I}(t_1)\hat{I}(t_2)}_c$ defined in the context of Eq.~\ref{eq:Pc0}, which turns out to be a
combination of the retarded and advanced response functions and the shot-noise correlator  \cite{kamenev,baym} for the current $\hat{I}(t)$,
is written as
\begin{align}
&\expect{\hat{I}(t_1)\hat{I}(t_2)}_c=Tr[I(t_1)G^>(t_1,t_2)I(t_2)G^<(t_2,t_1)]\label{eq:I2c},
\end{align}
where $I[\phi(t)]=\partial_\phi h[\phi]$.
The connected part of the power-spectrum of current fluctuations  can be computed using Eq.~\ref{eq:I2c} in Eq.~\ref{eq:Pc0}, which simplifies
after substituting Eq.~\ref{Gl}, to
\begin{align}
&P_c(\omega)=\sum_n\frac{2\Gamma  |\tilde{I}_{\lambda_1\lambda_2}(n)|^2}{\Gamma^2+(\omega-\tilde{\omega}_n)^2}\label{eq:Pc}
\end{align}
where $\tilde{\omega}_n=\omega-\epsilon_{\lambda_1}+\epsilon_{\lambda_2}+n\Omega_J$ and $\tilde{I}_{\lambda_1\lambda_2}(n)=\int_0^\tau dt e^{i n\Omega_J t}\expect{\varphi_{\lambda_1}(t)|I(t)|\varphi_{\lambda_2}(t)}$.
It is clear from the form of this equation that in the limit $2 e V\gg \Gamma$, one can expect the power-spectrum $P_c(\omega)$
to have peaks at the differences in quasi-energy $\omega\sim \epsilon_{\lambda_1}-\epsilon_{\lambda_2}$.
In the LZ limit where the voltage $V$ is much greater than the avoided crossing frequency so that $\theta\rightarrow \pi/2$ in Eq.~\ref{eq:epslambda},
 one can check from Eq.~\ref{eq:epslambda} the difference in eigenvalues
\begin{equation}
\epsilon_{1}-\epsilon_{0} \rightarrow \frac{2 e V}{2}
\end{equation}
giving rise to the fractional Josephson peaks in the current-noise spectrum.
This argument can be verified (as shown in Fig.~\ref{Fig2}) by numerically evaluating the power-spectrum in Eq.~\ref{eq:Pc} corresponding to the ABS spectrum in
Fig.~\ref{Fig1}. Interestingly, we find that in the intermediate regime between the fractional and conventional Josephson effect, the power-spectrum $P(\omega)$
can have peaks at incommensurate frequencies. Additionally, the terms proportional to $I_n$ in Eq.~\ref{eq:P} together with 
other ABSs in the conventional as well as TS junctions,
 which follow the conventional Josephson phase periodicity are expected to lead to the peaks
at the conventional Josephson frequencies. However, it is possible for the fractional Josephson contribution to dominate over the
conventional one as appears to be the case in recent experiments~\cite{rokhinson}.

\paragraph{Shapiro steps:}
The time-dependent current in a finite voltage biased SNS junctions can be detected more easily through the detection of Shapiro voltage steps
across the Josephson junction ~\cite{tinkham}. 
As discussed in the appendix, the Shapiro step experiment consists of injecting ac power into a resistively and capactively shunted
 SNS junction through the application of an 
ac voltage and measuring the response in the DC voltage across the SNS junction, which is biased to be in the over-damped regime (as in the ac fractional Josephson effect) so that it always has a finite DC voltage $V_{dc}\neq 0$ across it. 
 Therefore the Shapiro step experiment is the reciprocal measurement of the ac 
Josephson effect in essentially the same set-up (which is shown in Fig.~\ref{Fig3} of the appendix). 
In a conventional (low transparency) 
SNS junction where $V_{dc}$ is much smaller than the minimum gap in the junction, the application of an ac voltage with frequency $\omega$ 
leads to a plateau in the dc voltage across the SNS junction  as a function of $I_{tot}$ at $V_{dc}=\omega/2 e$. The range of $I_{tot}$  
over which this plateau exists is found to be proportional to the amplitude of the applied ac voltage $v_{ac,0}$ (as reviewed in the appendix).
For a finite temperature or otherwise noisy classical SNS junction, such as the one described by Eq.~\ref{eq:markov} earlier in the text, 
we expect that the SNS junction does not carry long time classical correlations. Therefore the response to $v_{ac,0}$ should be analytic in the 
$v_{ac,0}\rightarrow 0$ limit and also be time-translation invariant. As a result the response of $V_{dc}$ scales as $v_{ac,0}^2$, which is 
different from the plateau. To compare the small $v_{ac,0}$ response with the plateau it is convenient to define the function 
\begin{equation}
f(\bar{V}_{dc}=I_{tot}R_1)=\frac{ I_{tot}R_1-V_{dc}}{R_1},\label{eq:f}
\end{equation}
which in the case of a plateau shows a kink near specific values of $V_{dc}$ that resembles the derivative of the peaks in the 
power-spectrum of the ideal SNS 
junction in the limit of  small $\omega$ (see appendix for explanation).
 In the classical fractional junction case,  where the power-spectrum has a peak at $\omega=e V_{dc}$
(see Eq.~\ref{eq:mc}), we find a similar  
 plateau with width in current $I_{tot}$  proportional to  $v_{ac,0}$, whenever $v_{ac,0}$ is large compared to the noise induced broadening 
of the power-spectrum peak. In the small $v_{ac,0}$ limit we argue that this plateau broadens out into a kink (henceforth referred to as a 
Shapiro kink ) whose 
weight is proportional to $v_{ac,0}^2$ and whose profile resembles the derivative of the broadened power-spectrum peak.
The resulting shape for $f(V_{dc})/v_{ac,0}^2$ is related to the response of the noise correlator of the current to shifts of the phase
\begin{equation}
\frac{f(V_{dc})}{v_{ac,0}^2}=\int dt dt' dt_0 F(t-t_0,t-t')\frac{\delta\expect{I_{sc}(t)I_{sc}(t_0)}}{\delta \phi_{cl}(t')},\label{eq:IShap}
\end{equation}
which depends only on the linear response properties of the junction and $F(t-t_0,t-t')$ written out explicitly in the appendix (see Eq.~\ref{eq:Vdc2}) 
is a response kernel of the external $RC$ circuit. 
The resulting kink profile function $f$ calculated using Eq.~\ref{eq:IShap}, as seen from the inset in Fig.~\ref{Fig2},
shows a kink corresponding to the derivative of the dissipated power-spectrum 
from the ac fractional Josephson effect.

\paragraph{Conclusion:}
We have shown that high transparency
conventional SNS junctions with weakly avoided zero-crossings~\cite{beenakker} can show a peak at $\omega\sim \Omega_J/2$ in the
power-spectrum thus showing a fractional ac Josephson effect even at voltages $V\ll \Delta$. In the case of coherent dynamics of the SNS junction,
we find peaks in the radiation at a frequency between $\Omega_J/2$ and $\Omega_J$, which we conjecture could lead to Shapiro kinks at
incommensurate values. The fractional Josepshon
effect arises here as a result of LZ processes resulting from a break-down of adiabaticity near the avoided crossing and exists
for voltages that are larger than the gap $E_g$ in the Andreev spectrum.
In contrast the fractional Josephson effect in TS systems arises from
protected zero-energy crossings in the ABS spectrum ~\cite{kitaev,yakovenko} and therefore would theoretically exist up to
$V\rightarrow 0$. In practice, damping from fermionic baths restricts the observation of the fractional Josephson
effect to finite (and often large) voltages $V$. Therefore, the ac fractional Josephson effect at finite voltages, similar to zero-bias
conductance peaks can arise both arise in conventional superconducting systems as a result of accidental fine tuning.

We acknowledge enlightening discussions with  Chris Laumann, Baruch Horovitz and Sumanta Tewari in the course of this work.
We acknowledge Microsoft Station Q and the NSF for support.
JS  thanks the Harvard Quantum Optics Center for support.

\appendix
\section{Shapiro steps in terms of the radiated power-spectrum}

\subsection{Introduction}
\begin{figure}
\centering
\includegraphics[scale=0.3,angle=0]{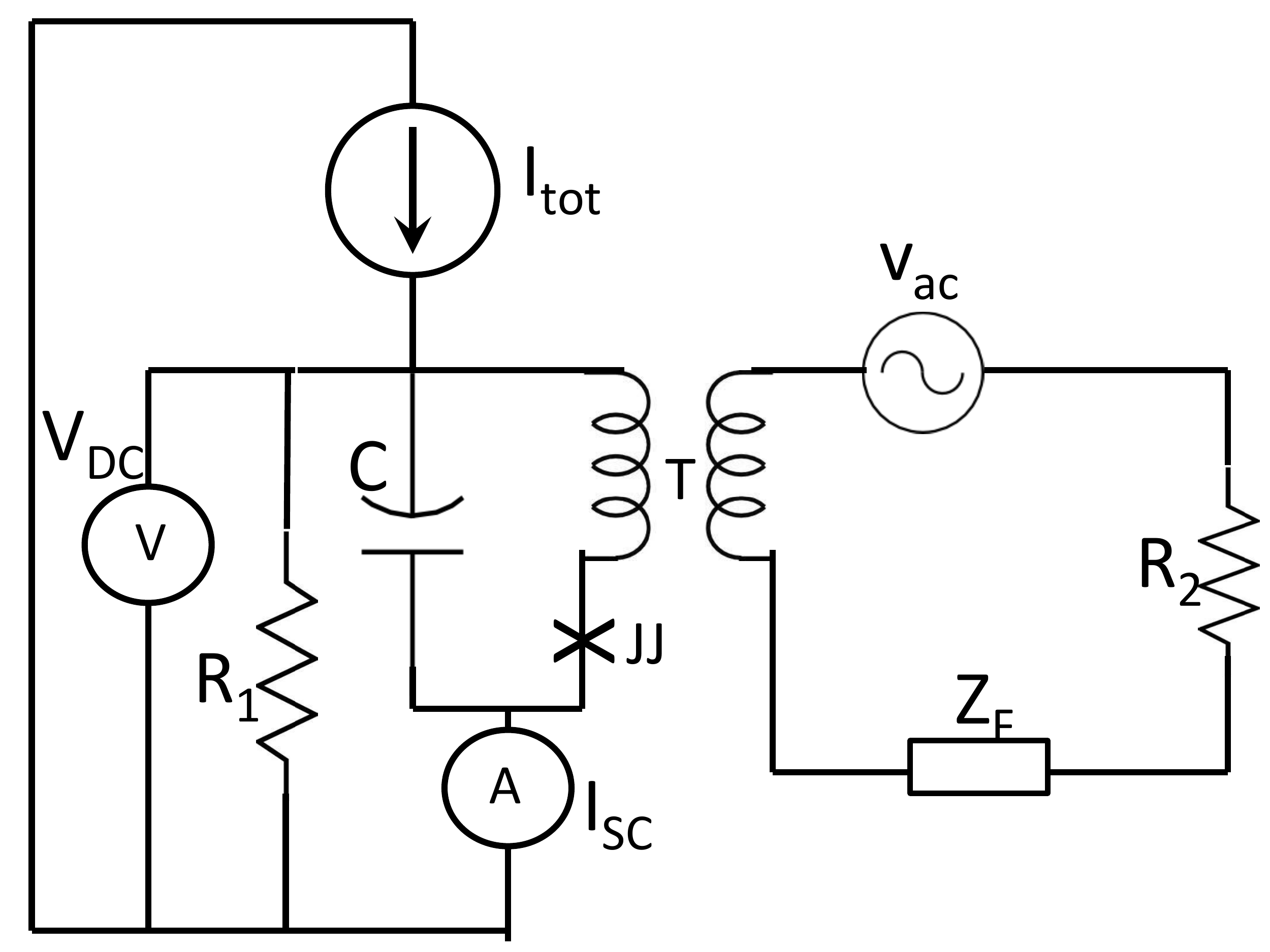}
\caption{Schematic circuit diagram for studying the ac Josephson effect and also the Shapiro steps across the Josephson junction $JJ$.
For the ac Josephson effect, we assume that the JJ is shunted by a vanishingly small parallel resistor $R_1$  and is biased by a large total current $I_{tot}\gg I_c$
 (where $I_c$ is the critical 
current of the JJ) so that the DC voltage across the JJ is $V_{dc}\approx I_{tot}/R_1$. The finite voltage $V_{dc}$ across the JJ creates an ac current through the $JJ$, which is 
mostly carried by the large capacitor $C$ and does not affect the voltage across $R_1$. This ac current is inductively coupled to the large resistor $R_2$ through the transformer $T$.
By choosing a filter $Z_F$ with a resonance at a frequency $\omega$, one can measure the power-spectrum of the current $\expect{I_{SC}(\omega)I_{SC}(-\omega)}$, which would 
have peaks dependent on $V_{dc}$. This is the ac Josephson effect. For the Shapiro step experiment, $R_1$ is chosen to be larger but still in the overdamped regime. An $ac$ voltage $v_{ac}(t)=v_{ac,0}\cos{\omega t}$ is then applied (after making the resistor $R_2$ vanishingly small). 
The dc voltage $V_{dc}$ when measured as a function of $I_{tot}$ shows steps or kinks which are referred to as Shapiro steps.
}\label{Fig3}
\end{figure}

In this section we consider the Shapiro step experiment and compare it with the ac Josephson effect using the circuit shown in Fig.~\ref{Fig3}.
The ac Josephson effect occurs when a finite DC voltage $V_{dc}$ is applied across the Josephson junction $JJ$ in Fig.~\ref{Fig3}, by setting 
the resistor $R_1\rightarrow 0$ and the current $I_{tot}$ correspondingly to $\infty$. These two limits are equivalent to applying a nearly 
ideal voltage source $V_{dc}=I_{tot}/R_1$. For the conventional Josephson junction with a current-phase relation
\begin{align}
I=I_c\sin{\phi},
\end{align}
the voltage $V_{DC}$ leads to a precession of the phase variable $\phi$ according to the equation
with $\phi(t)=4 e\pi V_{dc}t$.
This results in an ac current across the JJ 
\begin{equation}
I_J(t)=I_c\sin{(2 e V_{dc} t)},
\end{equation}
which is inductively coupled to the resistor $R_2$ through the transformer $T$. The power dissipated in the resistor $R_2$ as a function 
of the resonance frequency $\omega$ of the filter $Z_f$ shows a peak at $\omega = 2 e V_{dc}$, which is referred to as the 
ac Josephson effect.

The Shapiro step experiment is in a sense the reverse of the ac Josephson experiment. An ac current is now injected into the transformer $T$ in Fig.~\ref{Fig3} 
through the voltage source $v_{ac}=v_{ac,0}\cos{(\omega t)}$ and the DC voltage $V_{dc}$ is measured. To avoid externally imposing the voltage $V_{dc}$, one must now 
make the resistor $R_1$ finite (in contrast to being vanishingly small) but small enough so that JJ is in the over-damped regime.
 To avoid extra dissipation $R_2$ is made vanishingly small.
 The shunt capacitor $C$ is now 
important to avoid having extra ac voltage arising from the ac current in the $JJ$ flowing through the resistor $R_1$. The voltage $V_{dc}$ is now given by the equation
\begin{align}
V_{dc}=(I_{tot}-I_{SC})R_1.
\end{align}
In addition, the transformer (assuming it to be ideal and $1:1$) induces an ac voltage $-v_{ac,0}\cos{\omega t}$ across the JJ. The 
minus sign is a result of the fact that the transformer is in series with the junction and together these are shunted by a capacitor $C$, which forces 
the total ac voltage to be zero.
The total current through the JJ is given by 
\begin{align}
&I_{SC}(t)=I_c\sin{\{2 e V_{dc} t+\frac{v_{ac,0}}{\omega}\sin{(\omega t)}\}}\nonumber\\
&\approx I_c\sin{(2 e V_{dc} t)}-I_0\frac{v_{ac,0}}{2\omega}\cos{(2 e V_{dc}+\omega)t}\nonumber\\
&+I_c\frac{v_{ac,0}}{2\omega}\cos{(2 e V_{dc}-\omega)t}\label{eq:IJ}.
\end{align}
The last term in $I_{SC}(t)$ is the only low-frequency term, which can lead to a low-frequency ac component  $\delta v(t)$
in the voltage across the capacitor of the form
\begin{align}
&\delta v(t)=-I_c\frac{v_{ac,0}R_1}{2\omega\sqrt{1+(2 e V_{dc}-\omega)^2 R_1^2 C^2}}\nonumber\\
&\cos{\{(2 e V_{dc}-\omega)t+\bar{\phi}(\omega)\}}
\end{align}
where $\bar{\phi}(\omega)=\tan^{-1}{(R_1 C (2 e V_{dc}-\omega))}$.
This leads to an additional phase-fluctuation
\begin{align}
&\delta\phi(t)=-I_c R_1\frac{v_{ac,0}}{2\omega(2 e V_{dc}-\omega)\sqrt{1+(2 e V_{dc}-\omega)^2 R_1^2 C^2}}\nonumber\\
&\sin{\{(2 e V_{dc}-\omega)t+\bar{\phi}(\omega)\}}.
\end{align}
Because of the pole when $V_{dc}=\omega/2 e$, the above expression is only small and meaningful as a perturbation for $|2 e V_{dc}-\omega|>I_c R_1$. In this case, one can expand the low-frequency part of the expression for $I_{SC}(t)$ as 
\begin{align}
&\delta I_{SC}(t)\approx I_c\frac{v_{ac,0}}{2\omega}\cos{\{(2 e V_{dc}-\omega)t+\delta \phi(t)\}}\nonumber\\
&\approx I_c\frac{v_{ac,0}}{2\omega}\cos{\{(2 e V_{dc}-\omega)t\}}\nonumber\\
&+I_c^2 R_1\frac{v_{ac,0}^2}{4\omega^2 (2 e V_{dc}-\omega)\sqrt{1+(2 e V_{dc}-\omega)^2 R_1^2 C^2}}\\
&\sin{\{(2 e V_{dc}-\omega)t+\varphi(\omega)\}} \sin{\{2(2 e V_{dc}-\omega)t\}}\nonumber\\
&\approx I_c\frac{v_{ac,0}}{2\omega}\cos{\{(2 e V_{dc}-\omega)t\}}\nonumber\\
&-I_c^2 R_1\frac{v_{ac,0}^2}{4\omega^2 (2 e V_{dc}-\omega)}\cos{\{2(2 e V_{dc}-\omega)t+\varphi(\omega)\}}\nonumber\\
&+I_c^2 R_1\frac{v_{ac,0}^2}{4\omega^2 (2 e V_{dc}-\omega)\sqrt{1+(2 e V_{dc}-\omega)^2 R_1^2 C^2}}.
\end{align}
The dc voltage in this case is determined by the equation
\begin{equation}
I_{tot}R_1=V_{dc}+\frac{v_{ac,0}^2}{4\omega^2}\frac{(I_c R_1)^2}{(2 e V_{dc}-\omega)\sqrt{1+(2 e V_{dc}-\omega)^2 R_1^2 C^2}}.
\end{equation}
The above equation has no solution in the range 
\begin{equation}
|I_{tot}R_1-\frac{\omega}{2 e}|>\frac{v_{ac,0}(I_c R_1)}{2\omega}\sqrt{\frac{1}{2 e\sqrt{1+(2 e V_{dc}-\omega)^2 R_1^2 C^2}}},\label{eq:range1}
\end{equation}
which corresponds to voltage $V_{dc}$ very close to $\omega/2 e$, where the perturbation theory is expected to break down.

In the range where $V_{dc}\approx \omega/2 e$ the voltage readjusts the phase, so that $V_{dc}$ adjusts to $V_{dc}=\omega$. This
is the  state which produces the Shapiro step in the current $I_{tot}$\cite{tinkham}. In this case, 
the current in Eq.~\ref{eq:IJ} has a DC component given by
\begin{align}
&I_{SC,dc}(t)=I_c\frac{v_{ac,0}}{2\omega}\cos{\phi_{rel}(t)},\label{eq:Idc}
\end{align}
where $\phi_{rel}(t)$ is the slowly varying (or constant) relative phase between the superconducting phase and the ac perturbation.
If $\phi_{rel}(t)$ remains bounded in fluctuations i.e. $\phi_{rel}(t)\ll 2\pi$, then it is referred to as the locked state.
Since $I_{SC,dc}$ is bounded by $I_c\frac{v_{ac,0}}{2\omega}$ in magnitude the locked state can only be stable in the range 
\begin{equation}
\frac{\omega}{2 e R_1}-I_c\frac{v_{ac,0}}{2\omega}<I_{tot}<\frac{\omega}{2 e R_1}+I_c\frac{v_{ac,0}}{2\omega},\label{eq:range2}
\end{equation}
where $V_{dc}$ plateaus at the value $\omega$. 
The relative phase $\phi_{rel}(t)$ is now locked to 
\begin{align}
&\cos{\phi_{rel}(t)}=I_{tot}-\frac{\omega}{2 e R_1},\label{eq:phirel}
\end{align}
In this range the voltage $V_{dc}$ shows a plateau in the neighborhood $I_{tot}\approx \frac{\omega}{2 e}$ with width proportional to $v_{ac,0}$.
This range is approximately (i.e. overlaps only slightly) complimentary to the range in Eq.~\ref{eq:range1} where $V_{dc}$ has an 
oscillating in time component.

\subsection{Classical fractional Shapiro effect}
Let us now consider how a doubled voltage Shapiro step can exist in the classical SNS junction with an avoided crossing.
Using the classical Markov model in Eq.~\ref{eq:mc} ,we consider a current phase relation 
\begin{equation}
I_{SC}(t)= s(t)I_c\sin{(\phi(t)/2)},
\end{equation} 
where $s(t)=\pm 1$ is a state variable that randomly fluctuates on a characteristic time $\Gamma$.
If $s(t)$ is constant over all time,  based on the discussion in the previous section we expect that 
$V_{dc}$ is locked to $V_{dc}=\omega/ e R_1$ for a range of currents $I_{tot}\sim \omega/ e R_1$ which is given by Eq.~\ref{eq:range2}.
The DC current in the junction is now given by 
\begin{align}
&I_{SC,dc}(t)=I_c\frac{v_{ac,0}}{2\omega}s(t)\cos{\phi_{rel}(t)}.\label{eq:Idc1}
\end{align}

When $s(t)$ changes from $s(t)=+1$ to $s(t)=-1$, the DC current in Eq.~\ref{eq:Idc1} flips sign. This results in the DC voltage jumping to 
   $\frac{\omega}{ e}-2 I_c R_1\frac{v_{ac,0}}{2\omega}\cos{\phi_{rel}(t)}$. The DC voltage is now off resonance and leads to a
change of $\dot{\phi}_{rel}(t)\sim -2 I_c R_1\frac{v_{ac,0}}{2\omega}\cos{\phi_{rel}(t)}$.
Based on this equation, one expects the system to fall back in lock on a time-scale 
\begin{equation}
\tau=v_{ac,0}^{-1}\frac{2\omega}{I_c R_1}.
\end{equation}
If the $\Gamma$ i.e. the time between flips of $s(t)$ satisfies the constraint $\Gamma\gg \tau$, then one expects plateaus at $V_{dc}=\omega/e$ 
corresponding to the doubled voltage Shapiro steps. On the other hand, the limit of small $v_{ac}$ is behavior is somewhat different 
and one expects only kinks in the $V_{dc}$ versus $I_{tot}$ curves instead of true plateaus. This is because in the weak $v_{ac,0}$ limit, we 
expect the voltage $V_{dc}$ at a fixed current bias $I_{tot}$ to be analytic in $v_{ac,0}$ and also independent of the phase reference of the microwave.
Therefore we expect  $V_{dc}$ to vary with $v_{ac,0}$ with the second power $v_{ac,0}^2$.

 The presence of the step in
the voltage as a function of current $V_{dc}(I_{tot})$ 
leads to a divergence in the function  
\begin{equation}
f(\bar{V}_{dc}=I_{tot}R_1)=\frac{ I_{tot}R_1-V_{dc}}{R_1}.
\end{equation}
The slight subtlely in the abve definition, which is that $I_{tot}(V_{dc})$ is not a unique function of
  $V_{dc}(I_{tot})$ is technical and is avoided  whenever the 
plateau has a slight slope. 
In the small $v_{ac,0}$ limit we expect all quantities, including the function $f$ to be finite (corresponding to $I_{tot}$ as a function of $V_{dc}$ 
being smooth) and proportional to $f\propto v_{ac,0}^2$.
 The fact that at larger values of $v_{ac,0}$ a plateau in $V_{dc}(I_{tot})$(implying a divergent $f$) suggests that $f(V_{dc})$
 has a kink at the values of voltage where there is a plateau in the noiseless JJ limit.
 We will discuss how such a kink in the function $f$ is related to the peaks in the noise spectrum in the  $v_{ac,0}\rightarrow 0$ limit
 in the next section.

\subsection{Shapiro kinks in the noisy Josephson junction}
The small $v_{ac,0}$ limit in a noisy Josephson junction of the type considered in the previous section, in general does not give rise 
to a linear in $v_{ac,0}$ dc voltage plateau or a full phase lock. Instead it gives rise to  phase correlations that result in a 
kink with strength $v_{ac,0}^2$ as we discuss in this section. We will find that the Shapiro kink is a more universal feature 
of the Josephson effect and depends on fewer of the details of locking.

Let us consider the circuit in Fig.~\ref{Fig3} in the Shapiro step regime so that we can set $R_2=0$ and $R_1$ is finite. Let us define 
the integral
\begin{equation}
\phi(t)=\int_{-\infty}^t dt_1 (V_{dc}(t_1)-I_{tot}R_1)\label{eq:phi}
\end{equation}
as the phase fluctuation variable associated with the voltage difference across the capacitor.
 The correlators of the field $\phi(t)$ and hence $V_{dc}$ across $JJ$ can be calculated through a Keldysh action \cite{kamenev}
\begin{align}
&S[\phi_{cl},\phi_q]=S_0[\phi_{cl},\phi_q]\nonumber\\
&-2\int dt \phi_q[\frac{C}{2}\frac{d^2\phi_{cl}}{dt^2}+\frac{1}{2 R_1}\frac{d\phi_{cl}}{dt}+i \frac{T}{2 R_1}\phi_q]\nonumber\\
&+\frac{2 i}{R_1}\int dt dt' \frac{(\phi_q(t)-\phi_q(t'))^2}{2\sinh^2(\pi T(t-t'))/\pi T^2},
\end{align}
where $\phi_{cl}$ and $\phi_q$ are the classical and quantum parts of the fluctuating field $\phi$, $T$ is the temperature of the resistor, $S_0$ is 
the action of the $JJ$ after the ABSs have been integrated out.  
 We will now consider the high temperature limit where the quantum (or shot) noise (i.e. the last term) can be neglected
 compared to the Johnson-Nyquist noise and moreover, 
we will assume that $T$ is large enough so that $\phi_q$ is small and we can expand the action to lowest order in $\phi_q$.
 \begin{align}
&S[\phi_{cl},\phi_q]=-2\int dt \phi_q[\frac{C}{2}\frac{d^2\phi_{cl}}{dt^2}+\frac{1}{2 R_1}\frac{d\phi_{cl}}{dt}+\partial_{\phi_q} S_0[\phi_{cl},0]]\nonumber\\
&-(i\frac{T}{R_1}+\partial_{\phi_q} ^2 S_0[\phi_{cl},0])\phi_q^2.
\end{align}
Following Ref.~\cite{kamenev}, the correlators of the classical field $\phi_{cl}(t)$ can be obtained by integrating out $\phi_q$ and replacing it 
by a Langevin equation for $\phi_{cl}(t)$
\begin{align}
&\frac{C}{2}\frac{d^2\phi_{cl}}{dt^2}+\frac{1}{2 R_1}\frac{d\phi_{cl}}{dt}+I_{SC}[\phi_{cl}](t)=\xi(t)\nonumber\\
&-\frac{R_1}{2 T}\partial_{\phi_q} ^2 S_0[\phi_{cl},0](t,t')\xi(t')\label{eq:langevin}
\end{align}
where $\xi(t)$ is Gaussian correlated noise, which in the Nyquist noise dominated (i.e. high temperature) limit has a correlator 
\begin{equation}
\expect{\xi(t)\xi(t')}=\delta(t-t')\frac{T}{R_1}.
\end{equation} 
To obtain Eq.~\ref{eq:langevin} we have expanded to lowest order in $\partial_{\phi_q}^2S_0$, which is the noise spectrum of the current in the 
$JJ$ i.e. we have made the assumption that the noise from $JJ$ produces only a small voltage across $R$ compared to $V_{dc}$.
 The function $I_{SC}[\phi_{cl}](t)=\partial_{\phi_q} S_0[\phi_{cl},\phi_q]|_{\phi_q=0}$ 
is the expectation value of the supercurrent in the presence of $\phi_{cl}(t)$, which is expected to vanish in the dc limit.
Eq.~\ref{eq:langevin} is written more explicitly in terms of the noise correlator as 
\begin{align}
&\frac{C}{2}\frac{d^2\phi_{cl}}{dt^2}+\frac{1}{2 R_1}\frac{d\phi_{cl}}{dt}=\xi(t)\nonumber\\
&-\frac{R_1}{2 T}\int_{-\infty}^{t} dt'\expect{I(t)I(t')}_c[\phi_{cl}(t_1)]\xi(t'),\label{eq:langevin1}
\end{align}
where $\expect{I(t)I(t')}_c[\phi_{cl}(t_1)]$ is the connected part of the current-fluctuation correlator from  $JJ$.

Defining $\phi_{cl}^{(0)}(t)$ to be the solution
 to Eq.~\ref{eq:langevin} for $S_0=0$ i.e. 
\begin{equation}
\phi_{cl}^{(0)}(t)=\int_{-\infty}^t R dt'(1-e^{-\frac{(t-t')}{R C}})\xi(t'),\label{eq:phi0}
\end{equation}
  Eq.~\ref{eq:langevin1} may  be rewritten as 
\begin{align}
&\dot{\phi}_{cl}(t)-\dot{\phi}_{cl}^{(0)}(t)\nonumber\\
&=\frac{R_1}{2 T}\int_{-\infty}^{t} \frac{dt''}{C}e^{-\frac{(t-t'')}{RC}}\int_{-\infty}^{t''}dt' \expect{I(t'')I(t')}_c[\phi_{cl}(t_1)]\xi(t')\label{eq:langevin2}.
\end{align}
The mean dc voltage shift as a result of the junction $JJ$ is given by 
\begin{align}
&\delta V_{dc}(t)=\frac{R_1}{2 T}\int_{-\infty}^{t} \frac{dt''}{C}e^{-\frac{(t-t'')}{RC}}\int_{-\infty}^{t''}dt'\nonumber\\
& \expect{I(t'')I(t')}_c[\phi^{(0)}_{cl}(t_1)]\xi(t')\label{eq:Vdc}
\end{align}
In the small $R$ limit the variation of the phase is slow so that we can write 
\begin{align}
&\delta V_{dc}(t)=\frac{R_1}{2 T}\int_{-\infty}^{t} \frac{dt''}{C}e^{-\frac{(t-t'')}{RC}}\int_{-\infty}^{t''}dt'\nonumber\\
&\int \frac{d\phi}{2\pi} \frac{\delta\expect{I(t'')I(t')}}{\delta\phi^{(0)}_{cl}(t_1)}(\phi^{(0)}_{cl}(t_1)-\phi^{(0)}_{cl}(t'))\xi(t'),
\end{align}

Substituting Eq.~\ref{eq:phi0} and averaging over the Langevin noise $\xi(t)$ leads to 
\begin{align}
&\delta V_{dc}(t)=\frac{R_1}{2 C}\int_{t'<t_1<t''<t} dt''dt' dt_1 e^{-\frac{(t-t'')}{R_1 C}}(1-e^{-\frac{(t_1-t')}{R_1 C}})\nonumber\\
& \int \frac{d\phi}{2\pi} \frac{\delta\expect{I(t'')I(t')}_c}{\delta\phi^{(0)}_{cl}(t_1)}.\label{eq:Vdc2}
\end{align}
The integral over the phase $\phi=\phi_{cl}^{(0)}(t')$, which represents the part of the phase that diffuses because of the thermal 
noise from $R_1$, averages over all initial phases.
The shift in DC voltage $\delta V_{dc}=V_{dc}-I_{tot}R_1$ is directly proportional  to the Shapiro kink function $f$ defined in Eq.~\ref{eq:f}.

Eq.~\ref{eq:Vdc2}, which relates the shift of the voltage associated with a Shapiro kink 
to a derivative of the current-current correlator is the result of the derivation in this section.
Because of the integrals over time the DC voltage $V_{dc}$, contains only contributions from the low-frequency part of $I(t)$ in the 
current-current correlator, which are proportional 
to $v_{ac,0}$. Therefore $\delta V_{dc}(t)\propto v_{ac,0}^2$ as conjectured. Furthermore, the size of this response provides the 
size of the kink (apart from the factor of $v_{ac,0}^2$). Moreover we expects peaks in the noise spectrum, which is what is measured in 
the ac Josephson effect to correlate with the steps in the response of the noise.

\subsubsection{Computing the Shapiro kink}
The size of the Shapiro kinks are proportional to the derivative of the connected correlator $\expect{I(t)I(t')}_c$, which 
must also be expanded to second order in $v_{ac}$. Therefore the relevant low-frequency 
component of $ \frac{\delta\expect{I(t'')I(t')}}{\delta\phi^{(0)}_{cl}(t_1)}$ is a third order derivative of $\expect{I(t)I(t')}_c$ 
with respect to the phase. Considering the expression Eq.~\ref{eq:I2c}, we notice that terms where the derivative with respect to phase $\phi$
operates on the $U$ terms are suppressed by factors proportional to the critical current associated with the ABSs. Therefore, we can expect 
the dominant contribution to the derivative to come from the term where all of the derivatives with respect to $\phi$
 act on the phase-dependence of the current operator itself 
i.e.
\begin{align}
&\frac{\delta\expect{I(t'')I(t')}}{\delta\phi^{(0)}_{cl}(t_1)}\approx \frac{v_{ac}^2}{\omega^2}\delta(t''-t_1)\nonumber\\
&[\cos^2{\omega t''}\expect{\frac{\delta^3 I(t'')}{\delta \phi^3}I(t')}_c+\cos{\omega t''}\cos{\omega t'}\expect{\frac{\delta^2 I(t'')}{\delta \phi^2}\frac{\delta I(t')}{\delta\phi}}_c\nonumber\\
&+\cos^2{\omega t'}\expect{\frac{\delta I(t'')}{\delta \phi}\frac{\delta^2 I(t')}{\delta\phi^2}}_c]+A,
\end{align}
where $A$ contains the rest of the terms which have higher order correlators of the current $I$.

The second order in current $I$ contribution to $V_{dc}$ is thus written as 
\begin{align}
&\delta V_{dc}^{(2)}(t)=\frac{R_1}{2 C}\frac{v_{ac}^2}{\omega^2}\int_{t'<t''<t} dt''dt' e^{-\frac{(t-t'')}{R_1 C}}(1-e^{-\frac{(t''-t')}{R_1 C}})\nonumber\\
& \int \frac{d\phi}{2\pi}[\cos^2{\omega t''}\expect{I'''(t'')I(t')}_c+\cos{\omega t''}\cos{\omega t'}\expect{I''(t'') I'(t')}_c\nonumber\\
&+\cos^2{\omega t'}\expect{ I'(t'') I''(t')}_c],
\end{align}
where $I'(\phi)=\partial_{\phi} I(\phi)$,  $I''(\phi)=\partial_{\phi}^2 I(\phi)$ and  $I'''(\phi)=\partial_{\phi}^3 I(\phi)$.
Rewriting the products of the cosines as sums of cosines, we observe that since both $t'$ and $t''$ are integrated out, 
only the second term can have  a low-frequency contribution that is significant, 
so that 
\begin{align}
&\delta V_{dc}^{(2)}(t)\approx \frac{R_1}{2 C}\frac{v_{ac}^2}{\omega^2}\int_{t'<t''<t} dt''dt' e^{-\frac{(t-t'')}{R_1 C}}(1-e^{-\frac{(t''-t')}{R_1 C}})\nonumber\\
&\cos{\omega (t''-t')}\int \frac{d\phi}{2\pi}\expect{I''(t'') I'(t')}_c,
\end{align}
where the average over the overall phase $\phi$ is implicit in the remaining equations.

Following Eq.~\ref{eq:I2c}, the connected correlator for a pair of operators
 $\mathcal{O}^{(1)}(t)$ and $\mathcal{O}^{(2)}(t)$ is written as  
\begin{align}
&\expect{\mathcal{O}^{(1)}(t'')\mathcal{O}^{(2)}(t')}_c= \sum_{\lambda_1,\lambda_2}\mathcal{O}^{(1)}_{\lambda_1\lambda_2}(t'')\mathcal{O}^{(2)*}_{\lambda_1\lambda_2}(t')\nonumber\\
&e^{-\Gamma|t'-t''|}e^{i(\epsilon_{\lambda_1}-\epsilon_{\lambda_2})(t'-t'')},\label{eq:O2c}
\end{align}
where $\mathcal{O}^{(a)}_{\lambda_b\lambda_c}(t)=\expect{\varphi_{\lambda_b}(t)|\mathcal{O}^{(a)}(t)|\varphi_{\lambda_c}(t)}$ are $\tau$-periodic functions.
Assuming $\Gamma>1/(R_1 C)$ i.e. the broadening of the peaks appears is intrinsic, 
  \begin{align}
&\delta V_{dc}^{(2)}(t)\approx \frac{1}{2 C^2}\frac{v_{ac}^2}{\omega^2}\int_{t''<t} dt'' e^{-\frac{(t-t'')}{R_1 C}}\nonumber\\
&\partial_{\omega}\int_{-\infty}^0 dt' \sin{\omega t'}\int \frac{d\phi}{2\pi}\expect{I''(t'') I'(t'+t'')}_c.
\end{align}
Substituting the expression for the connected correlator from Eq.~\ref{eq:O2c}, leads to the result 
  \begin{align}
&\delta V_{dc}^{(2)}\approx \frac{R_1}{4 C}\frac{v_{ac}^2}{\omega^2}\sum_{n,\lambda_1,\lambda_2}\tilde{I}''_{\lambda_1,\lambda_2}(n)\tilde{I}'^*_{\lambda_1,\lambda_2}(n) \nonumber\\
&[\frac{1}{\{\omega-i\Gamma+(\epsilon_{\lambda_1}-\epsilon_{\lambda_2}+n\Omega_J)\}^2}\nonumber\\
&+\frac{1}{\{\omega+i\Gamma-(\epsilon_{\lambda_1}-\epsilon_{\lambda_2}+n\Omega_J)\}^2}],
\end{align}
where $\tilde{I}'_{\lambda_1\lambda_2}(n)=\int_0^\tau dt e^{i n\Omega_J t}\expect{\varphi_{\lambda_1}(t)|I'(t)|\varphi_{\lambda_2}(t)}$ and $\tilde{I}''_{\lambda_1\lambda_2}(n)=\int_0^\tau dt e^{i n\Omega_J t}\expect{\varphi_{\lambda_1}(t)|I''(t)|\varphi_{\lambda_2}(t)}$.
Note that averaging over $\phi$ has no consequences in the high T and low frequency limit used for this calculation.

The above dc voltage has peaks at the bias voltages $I_{tot}$, where the denominators $\epsilon_{\lambda_1}-\epsilon_{\lambda_2}+n\Omega_J-\omega$ vanish 
can be expected to show a kink structure.  However, the induced dc voltage $\delta V_{dc}$ has a profile which is related to the derivative of 
the peak-structure corresponding to the ac Josephson effect shown in Fig.~\ref{Fig2}. 

\end{document}